\documentclass[runningheads]{llncs}
\pdfoutput=1

\usepackage[T1]{fontenc}
\usepackage[utf8x]{inputenx}

\usepackage{cite}
\usepackage[dvipsnames]{xcolor}
\usepackage{graphicx}
\usepackage{xspace}
\usepackage[hyphens]{url}
\usepackage[breaklinks=true]{hyperref}

\usepackage{array}
\usepackage{enumitem}
\usepackage{amssymb}
\usepackage{amsmath}
\usepackage{cleveref}

\usepackage{tikz}
\usetikzlibrary{calc,automata,positioning,decorations.pathreplacing}
\tikzset{align at top/.style={baseline=(current bounding box.north)}}
\tikzstyle{every node}=[font=\scriptsize]
\tikzstyle{state} = [draw,fill=white,circle,thick,align=center,inner sep=0pt,minimum size=4.5mm]
\tikzstyle{lstate} = [draw,fill=white,rectangle,rounded corners,thick,align=center,inner sep=2pt]
\tikzstyle{dot} = [fill,circle,inner sep=0mm,minimum size=1.25mm,line width=0mm]

\Crefname{figure}{Fig.}{Figs.}
\crefname{figure}{fig.}{figs.}
\Crefname{tabular}{Tab.}{Tabs.}
\crefname{tabular}{tab.}{tabs.}
\Crefname{section}{Sect.}{Sects.}
\crefname{section}{sect.}{sects.}
\Crefname{equation}{Eq.}{Eqs.}
\crefname{equation}{eq.}{eqs.}
\creflabelformat{equation}{#2#1#3}

\setitemize{noitemsep,topsep=0pt,parsep=0pt,partopsep=0pt,leftmargin=12.0pt}
\setenumerate{noitemsep,topsep=0pt,parsep=0pt,partopsep=0pt,leftmargin=12.5pt}
\setdescription{noitemsep,topsep=0pt,parsep=0pt,partopsep=0pt,leftmargin=12.5pt}

\newsavebox{\foobox}
\newcommand{\slantbox}[2][.3]
  {%
    \mbox
      {%
        \sbox{\foobox}{#2}%
        \hskip\wd\foobox
        \pdfsave
        \pdfsetmatrix{1 0 #1 1}%
        \llap{\usebox{\foobox}}%
        \pdfrestore
      }%
  }

\newcommand{\modest}{\textsc{\mbox{Modest}}\xspace}
\newcommand{\toolset}{\textsc{\mbox{Modest} Toolset}\xspace}

\newcommand{\tool}[1]{\textsc{#1}}

\renewcommand{\max}{\ensuremath{\mathrm{max}}}
\renewcommand{\min}{\ensuremath{\mathrm{min}}}
\newcommand{\opt}{\ensuremath{\mathit{opt}}}
\newcommand{\mcsta}{\textsf{mcsta}\xspace}
\newcommand{\eg}{e.g.\ }
\newcommand{\ie}{i.e.\ }
\newcommand{\etal}{et al.\xspace}
\newcommand{\wrt}{w.r.t.\xspace}

\newcommand{\set}[1]{\ensuremath{\{\,#1\,\}}}
\newcommand{\tuple}[1]{\ensuremath{\langle #1 \rangle}}
\newcommand{\powerset}[1]{\ensuremath{2^{#1}}\xspace}
\newcommand{\defeq}{\mathrel{\vbox{\offinterlineskip\ialign{\hfil##\hfil\cr{\tiny \rm def}\cr\noalign{\kern0.30ex}$=$\cr}}}}

\newcommand{\QQ}{\ensuremath{\mathbb{Q}}\xspace}
\newcommand{\NN}{\ensuremath{\mathbb{N}}\xspace}

\newcommand{\True}[0]{\ensuremath{\mathit{true}}\xspace}
\newcommand{\False}[0]{\ensuremath{\mathit{false}}\xspace}
\newcommand{\Dist}[1]{\ensuremath{\mathit{Dist}({#1})}\xspace}
\newcommand{\DistQ}[1]{\ensuremath{\mathit{Dist}_{\hspace{-1pt}\scalebox{0.7}{\slantbox{$\mathbb{Q}$}}}({#1})}\xspace}
\newcommand{\support}[1]{\ensuremath{\mathit{spt}({#1})}\xspace}
\newcommand{\compactdots}{\makebox[1em][c]{.\hfil.\hfil.}}

\usepackage[vlined,linesnumbered]{algorithm2e}

\SetAlgoCaptionLayout{raggedright}
\SetAlCapFnt{\small}
\SetAlCapNameFnt{\small}
\SetAlCapSkip{\abovecaptionskip\relax}
\SetEndCharOfAlgoLine{}
\SetArgSty{textrm}
\SetKwInOut{Input}{Input}\SetKwInOut{Output}{Output}
\SetCommentSty{textit}
\SetKw{Break}{break}
\SetKwProg{Type}{type}{}{end}
\SetKwProg{Function}{function}{}{end}
\SetKwFor{ForEach}{foreach}{do}{}
\SetKwFor{WhileTrue}{repeat}{}{end}
\SetKw{Continue}{continue}
\SetKwRepeat{DoWhile}{repeat}{while}
\SetKwProg{Fn}{function}{}{}
\crefname{algocf}{alg.}{algs.}
\Crefname{algocf}{Alg.}{Algs.}

\makeatletter%
\g@addto@macro\normalsize{%
  \setlength\abovedisplayskip{3pt}%
  \setlength\belowdisplayskip{3pt}%
  \setlength\abovedisplayshortskip{-3pt}%
  \setlength\belowdisplayshortskip{3pt}%
}%
\makeatother

\makeatletter
\def\orcidID#1{\smash{\href{http://orcid.org/#1}{\protect\raisebox{-1.25pt}{\protect\includegraphics{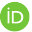}}}}}
\makeatother

\usepackage{pgfplots}
\pgfplotsset{compat=1.14}

\hypersetup{
  pdftitle = {Correct Probabilistic Model Checking with Floating-Point Arithmetic}
}

\begin{document}

\title{%
Correct Probabilistic Model Checking\\with Floating-Point Arithmetic%
\thanks{%
This work was supported by
NWO VENI grant no.\ 639.021.754.}%
}
\author{
Arnd Hartmanns\,\orcidID{0000-0003-3268-8674}
}

\institute{
University of Twente, Enschede, The Netherlands}
\maketitle

\begin{abstract}
Probabilistic model checking computes probabilities and expected values related to designated behaviours of interest in Markov models.
As a formal verification approach, it is applied to critical systems; thus we trust that probabilistic model checkers deliver correct results.
To achieve scalability and performance, however, these tools use finite-precision floating-point numbers to represent and calculate probabilities and other values.
As a consequence, their results are affected by rounding errors that may accumulate and interact in hard-to-predict ways.
In this paper, we show how to implement fast \emph{and} correct probabilistic model checking by exploiting the ability of current hardware to control the direction of rounding in floating-point calculations.
We outline the complications in achieving correct rounding from higher-level programming languages, describe our implementation as part of the \toolset's \mcsta model checker, and exemplify the tradeoffs between performance and correctness in an extensive experimental evaluation across different operating systems and CPU architectures.
\end{abstract}

\section{Introduction}
\label{sec:Introduction}

Given a Markov chain or Markov decision process (MDP~\cite{Put94}) model of a safety- or performance-critical system, probabilistic model checking (PMC) calculate quantitative properties of interest: the probability of (rare or catastrophic) failures, the expected recovery time after service interruption, or the long-run average throughput.
These properties involve probabilities or expected costs/rewards of sets of model behaviours, and are often specified in a temporal logic like PCTL~\cite{HJ94}.
As a formal verification approach, users place great trust in the results delivered by a PMC tool such as \tool{Prism}~\cite{KNP11}, \tool{Storm}~\cite{DJKV17}, ePMC~\cite{HLSTZ14}, or the \toolset's~\cite{HH14} \mcsta.
In contrast to classical model checkers for functional, Boolean-valued properties specified in \eg LTL or CTL~\cite{BK08}, a probabilistic model checker is inherently quantitative:
the input model contains real-valued probabilities and costs/rewards;
PCTL makes comparisons between real-valued constants and probabilities;
the most efficient algorithms numerically iterate towards a fixpoint;
and the final result itself may well be a real number.

In many cases, we can restrict to rational values, which simplifies the theory and facilitates ``exact'' algorithms operating on arbitrary-precision rational number datatypes.
These algorithms however only work for relatively small models (as shown in the most recent QComp 2020 competition of quantitative verification tools~\cite{BHKKPQTZ20}).
In this paper, we thus focus on the PMC techniques that scale to large problems:
those building upon iterative numerical algorithms, in particular value iteration (VI)~\cite{CH08}.
We also restrict to checking probabilistic reachability, \ie calculating the probability to eventually reach a goal state, as this is the core problem in PMC for MDP.
Embedded in the usual recursive algorithm for CTL, it allows us to check any (unbounded) PCTL property.

VI is a simple algorithm:
Starting from a trivial underapproximation of the reachability probability for each state of the model, it iteratively improves the value of each state based on the values of its successors.
The true reachability probabilities are the least fixpoint of this procedure, towards which the algorithm converges.
For roughly a decade, PMC tools implemented VI by stopping once the relative or absolute difference between subsequent iterations was below a threshold $\epsilon$.
Haddad and Monmege~\cite{HM14} showed in 2014\footnote{In fact, Wimmer \etal\cite{WKHB08} already in 2008 mention this problem in a more general setting, but unlike Haddad and Monmege's later work, they neither give a concrete counterexample nor propose a solution tailored to PMC.} that this in fact does not guarantee a difference of $\leq\epsilon$ between the reported and the true probability, putting in question the trust placed in PMC tools.
As a consequence, variants of VI were developed that provide \emph{sound}, \ie $\epsilon$-correct, results:
interval iteration (II)~\cite{HM18,BCCFKKPU14,BKLPW17},
sound value iteration (SVI)~\cite{QK18}, and
optimistic value iteration (OVI)~\cite{HK20}.
In this paper, we focus on II as the prototypical sound algorithm, but our approach can easily be transferred to SVI and OVI, too.
II additionally iterates on an overapproximation; its stopping criterion is simply the difference between over- and underapproximation being $\leq\epsilon$.

If all probabilities in an MDP are rational numbers, then the true reachability probability as well as all intermediate values in II are rational, too.
Yet implementing II with arbitrary-precision rationals is impractical since the smaller-and-smaller differences between intermediate values end up using excessive computation time and memory.
II is thus implemented with fixed-precision (usually 64-bit IEEE 754 \emph{double} precision) floating point numbers.
These, however, cannot represent all rationals, so operations must round to nearby representable values.
Although II is numerically benign, consisting only of multiplications and additions within $[0, 1]$, 
the default \emph{round to nearest, ties to even} policy can cause II to deliver incorrect results.
Wimmer \etal\cite{WKHB08} show an example where PMC tools incorrectly state that a simple PCTL property is satisfied by a small Markov chain due to the underlying numeric difference having disappeared in rounding.
We confirmed with current versions of \tool{Prism}, \tool{Storm}, and \mcsta that the problem persists to today, even when requesting a ``sound'' algorithm like II.
Wimmer \etal propose interval arithmetic to avoid such problems, cautioning that
\begin{quote}
\textit{[...] the memory consumption will roughly double, since two numbers for the interval bounds have to be stored [...].
The runtime will be higher by a small factor, because we need to derive lower and upper bounds for the intervals, requiring two model checking runs per sub-formula.}\hfill\cite[p.\,5]{WKHB08}
\end{quote}
They did not provide an implementation, and we are not aware of any to date.

\paragraph{Our contribution.}
We present the first PMC implementation that computes \emph{correct} lower and upper bounds on reachability probabilities despite using floating-point arithmetic.
We benefit from two developments since the suggestion of Wimmer \etal in 2008:
First, II (published in 2014) already works with intervals (albeit in a different way than Wimmer \etal likely envisioned) and necessarily doubles memory consumption compared to VI (as do SVI and OVI, so it appears an unavoidable cost of soundness).
In place of ``two model checking runs per sub-formula'', we can make the two interleaved computations inside II safe \wrt floating-point rounding.
Second, both hardware and programming language support for controlling the rounding direction in floating-point operations has improved, in particular with the AVX-512 instruction set in the newest x86-64 CPUs and widespread compiler support for C99's ``floating-point environment'' functions in the \texttt{fenv.h} header.
Nevertheless, it is nontrivial to achieve runtime that is only ``higher by a small factor''. 

The structure of this paper is as follows:
We first recap the essentials of PMC and the II algorithm (\Cref{sec:PMC}) as well as the problems and solutions related to rounding in fixed-precision floating-point arithmetic (\Cref{sec:Floats}).
We then present our new approach (\Cref{sec:CorrectII}), including important implementation aspects.
As we work on the core algorithms of PMC, the performance of our approach is crucial to its adoption in tools; we thus performed extensive experiments across different software and hardware configurations on models from the Quantitative Verification Benchmark Set (QVBS)~\cite{HKPQR19}, which we report in \Cref{sec:Experiments}.

\section{Probabilistic Model Checking}
\label{sec:PMC}

We write $\set{ x_1 \mapsto y_1, \dots }$ to denote the function that maps all $x_i$ to $y_i$. 
Given a set $S$, its powerset is $\powerset{S}$.
A (discrete) \emph{probability distribution} over $S$ is a function $\mu \in S \to [0, 1]$ with countable \emph{support} $\support{\mu} \defeq \set{ s \in S \mid \mu(s) > 0 }$ and $\sum_{s \in \support{\mu}} \mu(s) = 1$.
$\Dist{S}$ is the set of all probability distributions over $S$.
If $\mu(s) \in \QQ$ for all $s \in S$, \ie all probabilities are rational numbers, then we call $\mu$ a \emph{rational probability distribution}; the set of all of them is $\DistQ{S}$.

\paragraph{Markov decision processes}
(MDP)~\cite{Put94} combine the nondeterminism of Kripke structures with the finite random choices of discrete-time Markov chains (DTMC).

\begin{definition}
\label{def:MDP}
A \emph{Markov decision process} (MDP) is a triple
$M = \tuple{S, s_I, T}$
where
$S$ is a finite set of \emph{states} with \emph{initial state} $s_I \in S$ and
$T \colon \mathit{S} \to \powerset{\DistQ{S}}$ is the \emph{transition function}.
$T(s)$ must be finite and non-empty for all $s \in S$.
\end{definition}
For $s \in S$, an element $\mu$ of $T(s)$ is a \emph{transition}, and if $s' \in \support{\mu}$, then the transition has a \emph{branch} to successor state $s'$ with probability $\mu(s')$.
If $|T(s)| = 1$ for all $s \in S$, then $M$ is a DTMC.

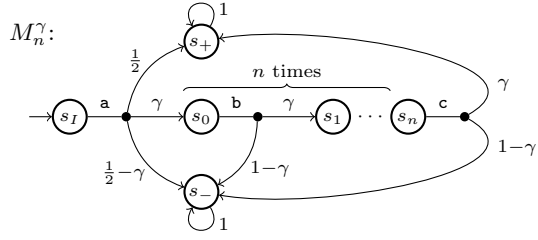
\begin{figure}[t]
\centering
\begin{tikzpicture}[on grid,auto]
  \node[state] (s0) {$\mathstrut s_I$};
  \coordinate[left=0.3 of s0.west] (start);
  \node[] (me) [above left=1.1 and 0.5 of s0] {\small$M_n^\gamma$:};
  \node[dot] (n0) [right=0.75 of s0] {};
  \node[state] (s1) [right=1.0 of n0] {$\mathstrut s_0$};
  \node[state] (sp) [above right=1.0 and 1.0 of n0] {$\mathstrut s_+$};
  \node[state] (sm) [below right=1.0 and 1.0 of n0] {$\mathstrut s_-$};
  \node[dot] (n1) [right=0.75 of s1] {};
  \node[state] (s2) [right=1.0 of n1] {$\mathstrut s_1$};
  \node[state] (sn) [right=1.0 of s2] {$\mathstrut s_n$};
  \node[] (ddd) [right=0.52 of s2] {$\cdots$};
  \node[dot] (nn) [right=0.75 of sn] {};
  ;
  \path[-]
    (s0) edge[] node[inner sep=3pt] {\texttt{a}} (n0)
    (s1) edge[] node[inner sep=3pt] {\texttt{b}} (n1)
    (sn) edge[] node[inner sep=3pt] {\texttt{c}} (nn)
  ;
  \path[->]
    (start) edge node {} (s0)
    (n0) edge [bend left] node[pos=0.67,left,inner sep=5pt] {$\frac{1}{2}$} (sp)
    (n0) edge [] node[above,inner sep=2pt] {$\gamma$} (s1)
    (n0) edge [bend right] node[pos=0.67,left,inner sep=4pt] {$\frac{1}{2}{-}\gamma$} (sm)
    (n1) edge [] node[above,inner sep=2pt] {$\gamma$} (s2)
    (n1) edge [bend left] node[pos=0.67,right,inner sep=4pt] {$1{-}\gamma$} (sm)
    (nn) edge [out=30,in=10] node[pos=0.2,right,inner sep=4pt] {$\gamma$} (sp)
    (nn) edge [out=-30,in=-10] node[pos=0.2,right,inner sep=4pt] {$1{-}\gamma$} (sm)
    (sm) edge [loop,out=-60,in=-120,looseness=5] node[right,pos=0.25,inner sep=2pt] {$1$} (sm)
    (sp) edge [loop,out=60,in=120,looseness=5] node[right,pos=0.25,inner sep=2pt] {$1$} (sp)
  ;
  \path[draw,decorate,decoration=brace]
    ([yshift=5pt,xshift=-2pt]s1.north west) -- ([yshift=5pt,xshift=-2pt]sn.north west) node[midway,above,yshift=2pt] {$n$ times}
  ;
\end{tikzpicture}
\caption{Example parametrised MDP $M_n^\gamma$}
\label{fig:ExampleMDP}
\end{figure}

\begin{example}
\label{ex:MDP}
\Cref{fig:ExampleMDP} shows our example MDP $M_n^\gamma$, which is actually a DTMC.
It is a simplified and parametrised version of the counterexample of Wimmer \etal\cite[Fig.~2]{WKHB08}.
It is parametrised in terms of $n \in \NN$ (determining the number of chained states with transitions labelled \texttt{b}) and $\gamma \in (0, 0.5)$ (changing some probabilities).
We draw transitions as lines to an intermediate node from which probability-labelled branches lead to successor states.
We omit the intermediate node for transitions with a single branch, and label some transitions to easily refer to them.
$M^e$ has $4+n$ states and transitions, and $7+2n$~branches.
\end{example}
In practice, higher-level modelling languages like \modest~\cite{HHHK13} are used to specify MDP.
The semantics of an MDP is captured by its \emph{paths}.
A path represents a concrete resolution of all nondeterministic and probabilistic choices.
Formally:%
%
%
%
\begin{definition}
A \emph{finite path} is a sequence
$\pi_\mathrm{fin} = s_0\, \mu_0\, s_1\, \mu_1 \dots  \mu_{n-1} s_n$
where $s_i \in S$ for all $i \in \set{ 0, \dots, n }$ and $\exists\,\mu \in T(s_i)\colon \mu(s_{i+1}) > 0$ for all $i \in \set{ 0, \dots, n - 1 }$.
Let $|\pi_\mathrm{fin}| \defeq n$ and $\mathrm{last}({\pi_\mathrm{fin}}) \defeq s_n$.
$\Pi_\mathit{fin}(s)$ is the set of all finite paths starting in $s$.
A \emph{path} is an analogous infinite sequence~$\pi$, and $\Pi(s)$ is the set of all paths starting in $s$.
We write $s \in \pi$ if $\exists\, i \colon s = s_i$. 
\end{definition}
A scheduler (or \emph{adversary}, \emph{policy} or \emph{strategy}) only resolves the nondeterministic choices of~$M$.
For this paper, memoryless deterministic schedulers suffice~\cite{BA95}.

\begin{definition}
\label{def:Scheduler}
A function $\mathfrak{s} \colon S \to \Dist{S}$ is a \emph{scheduler} if, for all $s \in S$, we have $\mathfrak{s}(s) \in T(s)$.
The set of all schedulers of~$M$ is $\mathfrak{S}(M)$.
\end{definition}
We are interested in \emph{reachability probabilities}.
Let $M|_\mathfrak{s} = \tuple{S, s_I, T|_\mathfrak{s}}$ with $T|_\mathfrak{s}(s) = \set{\mathfrak{s}(s)}$ be the DTMC induced by $\mathfrak{s}$ on $M$.
Via the standard cylinder set construction~\cite[Sect.\ 2.2]{FKNP11} on $M|_\mathfrak{s}$, a scheduler induces probability measures $\mathbb{P}_\mathfrak{s}^{M,s}$ on measurable sets of paths starting in $s \in S$.

\begin{definition}
For state $s$ and goal state $g \in S$, the maximum and minimum \emph{probability of reaching $\boldsymbol{g}$} from $s$ is defined as $\mathrm{P}_{\!\max}^{M,s}(\diamond\: g) = \sup_{\mathfrak{s} \in \mathfrak{S}} \mathbb{P}_\mathfrak{s}^{M,s}(\set{ \pi \in \Pi(s) \mid g \in \pi })$ and $\mathrm{P}_{\!\min}^{M,s}(\diamond\: g) = \inf_{\mathfrak{s} \in \mathfrak{S}} \mathbb{P}_\mathfrak{s}^{M,s}(\set{ \pi \in \Pi(s) \mid g \in \pi })$, respectively.
\end{definition}
The definition extends to sets $G$ of goal states.
We omit the superscript for $M$ when it is clear from the context, and if we omit that for $s$, then $s = s_I$.
From now on, whenever we have an MDP with a set of goal states $G$, we assume w.l.o.g.\ that all $g \in G$ are absorbing, \ie every $g$ only has one self-loop transition. 

\begin{definition}
\label{def:MEC}
A \emph{maximal end component} (MEC) of $M$ is a maximal \mbox{(sub-)}MDP $\tuple{S', T', s_I'}$ where $S' \subseteq S$, $T'(s) \subseteq T(s)$ for all $s \in S'$, and the directed graph with vertex set $S'$ and edge set $\set{ \tuple{s, s'} \mid \exists\,\mu \in T'(s) \colon \mu(s') > 0 }$ is strongly connected.
\end{definition}

\subsection{Algorithms}

\begin{algorithm}[t]
\Function{$\texttt{II}(M = \tuple{S, s_I, T}, G, \opt, \epsilon)$}{
  \smallskip
  \tcp{Preprocessing}
  \lIf(\tcp*[f]{collapse MECs}){$\opt = \max$}{%
    $M := \texttt{CollapseMECs}(M, G)$\label{alg:II:MECs}
  }
  $S_0 := \texttt{Prob0}(M, G, \opt)$, $S_1 := \texttt{Prob1}(M, G, \opt)$\label{alg:II:ZeroOneStates}\tcp*{identify 0/1 states}
  $l := \set{ s \mapsto 0 \mid s \in S \setminus S_1 } \cup \set{ s \mapsto 1 \mid s \in S_1 }$\label{alg:II:InitLower}\tcp*{initialise lower vector}
  $u := \set{ s \mapsto 0 \mid s \in S_0 } \cup \set{ s \mapsto 1 \mid s \in S \setminus S_0 }$\label{alg:II:InitUpper}\tcp*{initialise upper vector}
  \smallskip
  \tcp{Iteration}
  \While(\tcp*[f]{while relative error $> \epsilon$:}){$(u(s_I) - l(s_I))/l(s_I) > \epsilon$\label{alg:II:While}}{%
    \ForEach(\tcp*[f]{update non-0/1 states:}){$s \in S \setminus (S_0 \cup S_1)$\label{alg:II:Foreach}}{
      $l(s) := \opt_{\mu \in T(s)} \sum_{s' \in \support{\mu}}{\mu(s') \cdot l(s')}$\label{alg:II:Lower}\tcp*{iterate lower vector}
      $u(s) := \opt_{\mu \in T(s)} \sum_{s' \in \support{\mu}}{\mu(s') \cdot u(s')}$\label{alg:II:Upper}\tcp*{iterate upper vector}
    }
  }
  \Return{$\frac{1}{2}(u(s_I) - l(s_I))$}\label{alg:II:Return}
}%
\caption{Interval iteration for probabilistic reachability}
\label{alg:II}
\end{algorithm}

\paragraph{Interval iteration}\cite{HM14,BCCFKKPU14,BKLPW17,HM18}
computes reachability probabilities $p(s) = \mathrm{P}_{\opt}^{s}(\diamond\: G)$.
We show the basic algorithm as \Cref{alg:II}.
It iteratively refines vectors $l$ and $u$ that map each state to a value in $\QQ$ such that, at all times, we have $l(s) \leq p(s) \leq u(s)$.
In each iteration, the values in $l$ and $u$ are updated for all relevant states (line~\ref{alg:II:Foreach}) via the classic Bellman equations of value iteration (lines \ref{alg:II:Lower}-\ref{alg:II:Upper}). 
Their least fixpoint is $p$, towards which $l$ converges from below.
Some preprocessing is needed to ensure that the fixpoint is unique and also $u$ converges towards $p$:
for maximisation, we need to collapse MECs into single states (line~\ref{alg:II:MECs}).
This can be be done via graph-based algorithms (see \eg\cite{CH11}) that only consider the graph structure of the MDP as in \Cref{def:MDP} but do not perform calculations with the concrete probability values.
For both maximisation and minimisation, we need to identify the sets $S_0$ and $S_1$ such that $\forall s \in S_0\colon p(s) = 0$ and $\forall s \in S_1\colon p(s) = S_1$ (line~\ref{alg:II:ZeroOneStates}).
This can equally done via graph-based algorithms~\cite[Algs.~1-4]{FKNP11}.
We then initialise $l$ and $u$ to trivial under-/overapproximations of $p$ (lines \ref{alg:II:InitLower}-\ref{alg:II:InitUpper}).
Iteration stops when the relative difference between $l$ and $u$ at $s_I$ is at most $\epsilon$ (which is often chosen as $10^{-3}$ or $10^{-6}$).
The corresponding check in line~\ref{alg:II:While} assumes that division by zero results in $+\infty$, as is the default in IEEE 754.
By convergence of $l$ and $u$ towards the fixpoint, II terminates, and we eventually return a value $\hat p$ with the guarantee that $p(s_I) \in [ (1 - \epsilon) \cdot \hat p, (1 + \epsilon) \cdot \hat p ]$.
This makes II \emph{sound}.

\paragraph{PCTL.}
The temporal logic PCTL~\cite{HJ94} allows us to construct complex branching-time properties.
It takes standard CTL~\cite{BK08} and replaces the $\texttt{A}(\psi)$ (``for all paths $psi$ holds'') and $\texttt{E}(\psi)$ (``there exists a path for which $\psi$ holds'') operators by the probabilistic operator $\texttt{P}_{\sim c}(\psi)$ for ``under all schedulers, the probability of the measurable set of paths for which $\psi$ holds is $\sim c$'' where ${\sim} \in \set{ <, \leq, >, \geq }$ and $c \in [0, 1]$.
To model-check a PCTL formula on MDP $M$, we follow the standard recursive CTL model checking algorithm~\cite[Sect.~6.4]{BK08} except for the \texttt{P} operator, which can be reduced to computing reachability probabilities.
For the ``finally''/``eventually'' case $\texttt{P}_{\sim c}(\texttt{F}\,\phi)$, we can directly use interval iteration:
Let $S_\phi$ be the set of states recursively determined to satisfy $\phi$.
Call $\texttt{II}(M, S_\phi, \opt_\sim, \epsilon)$ of \Cref{alg:II} with $\opt_\sim = \max$ if ${\sim} \in \set{ <, \leq }$ and $\opt_\sim = \min$ otherwise, with two modifications:
Change the stopping criterion of line~\ref{alg:II:While} to check the difference for \emph{all} states, and in line~\ref{alg:II:Return}, return the set $S_\texttt{P} \defeq \set{ s \in S \mid \forall x \in [l(s), u(s)] \colon x \sim c }$.
If $\exists s \in \S, x \in [l(s), u(s)] \colon x \nsim c$, however, we would need to either abort and report an ``unknown'' situation, or continue with a reduced $\epsilon$ until we can (hopefully eventually) decide the comparison.
None of \tool{Prism}, \tool{Storm}, and \mcsta appear to perform this extra check, though.
In this paper, we only use PCTL for non-nested top-level $\texttt{P}_*(\texttt{F} \ldots)$ operators; the results are then \True if $s_I \in S_\texttt{P}$, should be $\mathit{unknown}$ in case the ``unknown'' situation applies to $s_I$, and are \False otherwise.

\section{Floating-Point Arithmetic}
\label{sec:Floats}

All current implementations of II that we are aware of (in \tool{Prism}, \tool{Storm}, and \mcsta) use IEEE 754 double-precision floating-point arithmetic to represent (a)~the probabilities of the branches in the MDP and (b)~the values in $l$ and $u$ during iteration.
A floating-point number is stored as a \emph{significant} $d$ and an \emph{exponent} $e$ \wrt to an agreed-upon \emph{base} $b$ such that it represents the value $d \cdot b^e$.
We fix $b = 2$.
IEEE 754 \emph{double} precision uses 64 bits in total, of which 1 is a sign bit, 52 are for $d$, and 11 are for $e$.
Standard alternatives are 32-bit \emph{single} precision (1 sign, 23 bits for $d$, and 8 for $e$) and the 80-bit x87 \emph{extended} precision format (with 1 sign bit, 64 for $d$, and 15 for $e$).
The subset of \QQ that can be represented in such a representation is determined by the numbers of bits for $d$ and $e$.
For example, $\frac{1}{2}$ or $\frac{7}{8}$ can be represented exactly in all formats, but $\frac{1}{10}$ cannot.
IEEE 754 prescribes that all basic operations (addition, multiplication, etc.) are performed at ``infinite precision'' with the result \emph{rounded} to a representable number.
The default rounding mode is to round to the nearest such number, choosing an even value in case of ties (\textit{round to nearest, ties to even}).
In single precision, $\frac{1}{10}$ is thus by default rounded to
$$13421773 \cdot 2^{-27} = 0.100000001490116119384765625.$$
A single rounded operation leads to an error of at most distance between the two nearest representable numbers.
In iterative computations, however, rounding may happen at every step. 
A striking example of the consequences 
is the failure of an American Patriot missile battery to intercept an incoming Iraqi Scud missile in February 1992 in Dharan, Saudi Arabia~\cite{GAO92}, which resulted in 28 fatalities.
The Patriot system calculated time in seconds by multiplying its internal clock's value by a rounded binary representation of $\frac{1}{10}$. 
After 100 hours of continuous operation, this lead to a cumulative rounding error large enough to miscalculate the incoming missile's position by more than half a kilometre~\cite{Arn00}.

\subsection{Errors in Probabilistic Model Checking}
\label{sec:ErrorsInPMC}

In II, we iteratively accumulate and multiply rounded floating-point values in the $l$ and $u$ vectors with potentially already-rounded values representing the rational probabilities of the model's branches.
Using the default rounding mode, how can we be sure that the final result does not miss the true probability by more than half a kilometre, too?

Following Wimmer \etal\cite{WKHB08}, let us consider MDP $M_n^\gamma$ of \Cref{fig:ExampleMDP} again, and determine whether $\texttt{P}_{\leq \frac{1}{2}}(\diamond\: \set{ s_+ })$ holds.
The model is acyclic, so it is easy to see that
$$p \defeq \mathrm{P}_{\max}(\diamond\: \set{ s_+ }) = \frac{1}{2} + \gamma^{n + 2} > \frac{1}{2}.$$
Let us fix $n = 1$ and $\gamma = 10^{-6}$.
Then $p = \frac{1}{2} + 10^{-18}$.
This value cannot be represented in double precision, and is by default rounded to $0.5$.

We have encoded $M_n^\gamma$ in the \modest and \tool{Prism} languages, and checked the answers returned by \tool{Prism} 4.7, \tool{Storm} 1.6.4, and \mcsta 3.1 for the property.
The correct result would be \False.
\tool{Prism} returns \True in its default configuration, which uses an unsound algorithm, and \False when requesting an algorithm with exact rational arithmetic, for which $M_n^\gamma$ is small enough.
If we explicitly request \tool{Prism} to use II, then the result depends on the specified $\epsilon$:
for $\epsilon \geq 10^{-11}$, we get the correct result of \False; for \emph{smaller} $\epsilon \leq 10^{-12}$, \ie \emph{higher precision}, however, we incorrectly get \True.
\tool{Storm} incorrectly returns \True in its default configuration as well as when we request a sound algorithm via the \texttt{-{}-sound} parameter.
Only when using an exact rational algorithm via the \texttt{-{}-exact} parameter does \tool{Storm} correctly return \False.
\mcsta, when using II (\texttt{-{}-alg IntervalIteration}), incorrectly returns \True, and additionally reports that it computed $[l(s_I), u(s_I)]$ as $[0.5, 0.5]$, thus not including the true value of~$p$.
Other algorithms are not immune to the problem, either; for example, \mcsta also answers \True when using SVI, OVI, and when solving the MDP as a linear programming problem via the Google OR Tools' GLOP LP solver.

This example shows that using a sound algorithm does not guarantee correct results.
The problem is not specific to cases of small probabilities like $\gamma = 10^{-6}$ in the MDP; we can achieve the same effect using arbitrarily higher values of $\gamma$ if we just increase $n$ a little.
Such bounded try-and-retry chains where ``normal'' probabilities in the model result in very small values during iteration and on the final result are not uncommon in the systems often modelled as MDPs, \eg backoff schemes in communication protocols and randomised algorithms.
In general, tiny differences in probabilities in one place may result in significant changes of the overall reachability probability; for example, in two-dimensional random walks, the long-run behaviour when the probabilities to move forward or backward are both $\frac{1}{2}$ is vastly different from if they are $\frac{1}{2}+\delta$ and $\frac{1}{2}-\delta$, respectively, for any $\delta > 0$.

\subsection{On Precision and Rounding Modes}


In our concrete example, we may be able to avoid the problem by increasing precision:
In the 80-bit extended format supported by all x86-64 CPUs, $\frac{1}{2} + 10^{-18}$ is by default rounded to $5.000000000000000009\compactdots$, so there is a chance of obtaining \False unless other rounding during iterations would lose all the difference.
Extended precision is used for C's \texttt{long} \texttt{double} type by \eg the GCC compiler; it is thus readily accessible to programmers.
It is, however, the most precise format supported in common CPUs today; if we need more precision, we would have to resort to much slower software implementations using \eg the \href{https://www.mpfr.org/}{GNU MPFR library}.
Any a-priori fixed precision, however, just shifts the problem to smaller differences, but does not eliminate it.

The more general solution that we propose in this paper is to control the rounding mode of the floating-point operations performed in the II algorithm.
In addition to the default \textit{round to nearest, ties to even} mode, the IEEE 754 standard defines three \emph{directed} rounding modes:
\textit{round towards zero} (\ie truncation),
\textit{round towards $+\infty$} (\ie always round up), and
\textit{round towards $-\infty$} (\ie always round down).
As we will explain in \Cref{sec:CorrectII}, using the latter gives us an easy way to make the computations inside II \emph{safe}, \ie guarantee the under- and overapproximation invariants for $l$ and $u$, respectively.
Control of the floating-point rounding mode however appears to be a rarely-used feature of IEEE 754 implementations; consequently the level and style of support for it in CPUs and high-level programming languages is diverse.

\subsection{CPU Support for Rounding Modes}

\tool{Storm} and \mcsta run exclusively on x86-64 systems (with the upcoming ARM-based systems so far only supported via their x86-64 emulation layers), while \tool{Prism} additionally supports several other platforms via manual compilation.
Thus we focus on x64-64 in this paper as the platform probabilistic model checkers overwhelmingly run on today.

\paragraph{X87 and SSE.}
All x64-64 CPUs support two instruction sets to perform floating-point operations in double precision:
The x87 instruction set, originating from the 8087 floating-point coprocessor, and the SSE instruction set, which includes support for double precision since the Pentium~4's SSE2 extension.
Both implement operations according to the IEEE 754 standard.
Aside from architectural particularities such as its stack-based approach to managing registers, the x87 instruction set notably includes support for 80-bit extended precision.
In fact, by default, it performs all calculations in that extended precision, only rounding to double or single precision when storing values back to 64- or 32-bit memory locations.
This has the advantage of reducing the error across sequences of operations, but for high-level languages makes the results depend on the compiler's choices of when to load/store intermediate values in memory vs.\ keeping them in x87 registers.
The SSE instructions only support single and double precision.

Both the x87 and SSE instruction sets support all four rounding modes mentioned above.
The rounding mode of operations for x87 and SSE is determined by the current value of the x87 FPU control word stored in the x87 FPU control register or the current value of the SSE MXCSR control register, respectively.
That is, to change rounding mode, we need to obtain the current control register value, change the two bits determining rounding mode (with the other bits controlling other aspects of floating-point operations such as the treatment of NaNs), and apply the new value.
This is done via the FNSTCW/FLDCW instruction pair on x87, and VSTMXCSR/VLDMXCSR for SSE.
Rounding mode is thus part of the global (per-thread) state, and we must be careful to restore its original configuration when returning to code that does not expect rounding mode changes.
Frequent changes of rounding mode thus incur a performance overhead due to the extra instructions that must be executed for every change and their effects on \eg pipelining.

\paragraph{AVX-512.}
AVX-512 is the extension to 512 bits of the sequence of \emph{single instruction, multiple data} (SIMD) instruction sets in x84-64 processors that started with SSE.
It became available for general-purpose systems in high-end desktop (Skylake-X) and server (Xeon) CPUs in 2017, but it took until the 10th generation of Intel's Core mobile CPUs in 2019 before it was more widely available in end-user systems.
It is supposed to appear in AMD CPUs with the upcoming Zen 4 architecture.
Aside from its 512-bit SIMD instructions, AVX-512 crucially also includes new instructions for single floating-point values where the operation's rounding mode is specified as part of the instruction itself via the new ``EVEX'' encoding.
Of particular note for implementing II are the new VFMADD$(r_1 r_2 r_3)$SD \emph{fused multiply-add instructions} (the $r_i$ determining how the operand registers are used) that can directly be used for the sums of products in the Bellman equations in lines \ref{alg:II:Lower}-\ref{alg:II:Upper} of \Cref{alg:II}.
Overall, AVX-512 thus makes rounding mode independent of global state, and may improve performance by removing the need for extra instruction sequences to change rounding mode.

\subsection{Rounding Modes in Programming Languages}
\label{sec:RoundingInLanguages}

Support for non-default rounding modes is lacking in most high-level programming languages.
Java, C\#, and Python, for example, do not support them at all.
If II is implemented in such a language, there is consequently no hope for a high-performance solution to the rounding problems described earlier.

For C and C++, the C99 and C++11 standards introduced access to the \emph{floating-point environment}.
The \texttt{fenv.h}/\texttt{cfenv} headers include the \texttt{fegetround} and \texttt{fesetround} functions to query the current rounding mode and change it, respectively.
Implementations of these functions on x86-64 read/change both the x87 and SSE control registers accordingly.
In the remainder of this paper, we focus on a C implementation, but most statements hold for C++ analogously.
The level of support for the C99 floating-point features varies significantly between compilers; it is in particular still incomplete in Clang\footnote{The \href{https://clang.llvm.org/docs/UsersManual.html\#c}{documentation} as of October 2021 states that C99 support in Clang ``is feature-complete except for the C99 floating-point pragmas''.} and GCC~\cite[Further notes]{C99GCC}.
Still, both compilers provide access to the \texttt{fegetround}/\texttt{fesetround} functions (via the associated standard libraries), but GCC in particular is not rounding mode-aware in optimisations.
This means that, for example, subexpressions that are evaluated twice, with a change in rounding mode in between, may be compiled by GCC into a single evaluation before the change, with the resulting value stored in a register and reused after the rounding mode change.
This can even happen when using the \texttt{-frounding-math} option\footnote{The \href{https://gcc.gnu.org/onlinedocs/gcc/Optimize-Options.html}{documentation} as of Oct.\ 2021 states that \texttt{-frounding-math} ``does not currently guarantee to disable all GCC optimizations that are affected by rounding mode.''}.
Programmers thus need to inspect the generated assembly to ensure that no problematic transformations have been made, or try to make them impossible by declaring values \texttt{volatile} or inserting inline assembly ``barriers''.


Overall, C thus provides a standardised way to change x87/SSE rounding mode, but programmers need to be aware of compiler quirks when using these facilities.
Support for AVX-512 instructions that include rounding mode bits in C, on the other hand, is only slightly more convenient than programming in assembly as
we can use the intrinsics in the \texttt{immintrin.h} header; there is no standard higher-level abstraction of this feature in either C or C++.

\section{Correctly Rounding Interval Iteration}
\label{sec:CorrectII}

Let us now change II as in \Cref{alg:II} to consistently round in safe directions at every numeric operation.
Given that we can change or specify the rounding mode of all basic floating-point operations on current hardware, we expect that a high-performance implementation can be achieved.
First, the preprocessing steps require no changes as they are purely graph-based.
The changes to the iteration part of the algorithm are straightforward:
In line~\ref{alg:II:While},
$$\text{\lWhile{$(u(s_I) - l(s_I))/l(s_I) > \epsilon$}{$\ldots$},}$$
we round the results of the subtraction and of the division towards $+\infty$ to avoid stopping too early.
In line~\ref{alg:II:Lower},
$$\textstyle{l(s) := \opt_{\mu \in T(s)} \sum_{s' \in \support{\mu}}{\mu(s') \cdot l(s')}},$$
the multiplications and additions round towards $-\infty$ while the corresponding operations on the upper bound in line~\ref{alg:II:Upper} round towards $+\infty$.
Recall that all probabilities in the MDP are rational numbers, \ie representable as $\frac{\mathit{num}}{\mathit{den}}$ with $\mathit{num}, \mathit{den} \in \NN$.
We assume that $\mathit{num}$ and $\mathit{den}$ can be represented exactly in the implementation.
Then, in line~\ref{alg:II:Lower}, we calculate the floating-point values for the $\mu(s') = \mathit{num}/\mathit{den}$ by rounding towards $-\infty$.
In line~\ref{alg:II:Upper}, we round the result of the corresponding division towards $+\infty$.
Finally, instead of returning the middle of the interval in line~\ref{alg:II:Return}, we return $[l(s_I), u(s_I)]$ so as not to lose any information (\eg in case the result is compared to a constant as in the example of \Cref{sec:ErrorsInPMC}).

With these changes, we obtain an interval guaranteed to contain the true reachability probability if the algorithm terminates.
However, rounding away from the theoretical fixpoint in the updates of $l$ and $u$ means that we may reach an effective fixpoint---where $l$ and $u$ no longer change because all newly computed values round down/up to the values from the previous iteration---at a point where the relative difference of $l(s_I)$ and $u(s_I)$ is still above~$\epsilon$.
This will happen in practice:
In QComp 2020~\cite{BHKKPQTZ20}, \mcsta participated in the \emph{floating-point correct track} by letting VI run until it reached a fixpoint under the default rounding mode with double precision.
In 9 of the 44 benchmark instances that \mcsta attempted to solve in this way, the difference between this fixpoint and the true value was more than the specified $\epsilon$.
With safe rounding away from the true fixpoint, this would likely have happened in even more cases.

To ensure termination, we thus need to make one further change to the II of \Cref{alg:II}:
In each iteration of the \textbf{while} loop, we additionally keep track of whether any of the updates to $l$ and $u$ changes the previous value.
If not, we end the loop and return the current interval, which will be wider than the requested $\epsilon$ relative difference.
We refer to II will all of the these modifications as \emph{safely rounding interleaved II} (SR-III) in the remainder of this paper.

\subsection{Sequential Interval Iteration}

\begin{algorithm}[t]
\Function{$\texttt{SR-SII}(M = \tuple{S, s_I, T}, G, \opt, \epsilon)$}{
  $\ldots(\textit{preprocessing as in \Cref{alg:II}})\ldots$\;
  \Repeat{$\neg\mathit{chg} \vee (u(s_I) - l(s_I))/l(s_I) \leq \epsilon$\label{alg:SII:While}}{%
    $\mathit{chg} := \False$\;
    $\texttt{fesetround}(\textit{towards }{-}\infty)$\label{alg:SII:RoundLower}\;
    \ForEach{$s \in S \setminus (S_0 \cup S_1)$\label{alg:SII:ForeachLower}}{
      $l_\mathit{new} := \opt_{\mu \in T(s)} \sum_{s' \in \support{\mu}}{\mu(s') \cdot l(s')}$\label{alg:SII:Lower}\tcp*{iterate lower vector}
      \lIf{$l_\mathit{new} \neq l(s)$}{%
        $\mathit{chg} := \True$
      }
      $l(s) := l_\mathit{new}$\label{alg:SII:StoreLower}
    }
    $\texttt{fesetround}(\textit{towards }{+}\infty)$\label{alg:SII:RoundUpper}\;
    \ForEach{$s \in S \setminus (S_0 \cup S_1)$\label{alg:SII:ForeachUpper}}{
      $u_\mathit{new} := \opt_{\mu \in T(s)} \sum_{s' \in \support{\mu}}{\mu(s') \cdot u(s')}$\label{alg:SII:Upper}\tcp*{iterate upper vector}
      \lIf{$u_\mathit{new} \neq u(s)$}{%
        $\mathit{chg} := \True$
      }
      $u(s) := u_\mathit{new}$\label{alg:SII:StoreUpper}
    }
  }
  \Return{$[l(s_I), l(s_I)]$}\label{alg:SII:Return}
}%
\caption{Safely rounding sequential interval iteration for x87 or SSE}%
\label{alg:SII}
\end{algorithm}

When using the x87 or SSE instruction sets to implement SR-III, we need to insert a call to \texttt{fesetround} just before line~\ref{alg:II:Lower}, and another just before line~\ref{alg:II:Upper}.
If, for an MDP with $n$ states, we need $m$ iterations of the \textbf{while} loop, we will make $2 \cdot n \cdot m$ calls to \texttt{fesetround}.
This might significantly impact performance for models with many states, or that need many iterations (such as the \textit{haddad-monmege} model of the QVBS, which requires $7$~million iterations with $\epsilon = 10^{-6}$ despite only having $41$ states).
As an alternative, we can rearrange the iteration phase of II as shown in \Cref{alg:SII}:
We first update $l$ for all states (lines \ref{alg:SII:ForeachLower}-\ref{alg:SII:StoreLower}), then $u$ for all states (lines \ref{alg:SII:ForeachUpper}-\ref{alg:SII:StoreUpper}), with the rounding mode changes in between (lines \ref{alg:SII:RoundLower} and~\ref{alg:SII:RoundUpper}).
We call this variant of II \emph{safely rounding sequential II} (SR-SII).
It only needs $2 \cdot m$ calls to \texttt{fesetround}, which should improve its performance.
However, it also changes the memory access pattern of II with an a priori unknown effect on performance.

\subsection{Implementation Aspects}

We have implemented III, SII, SR-III, and SR-SII in \mcsta.
While \mcsta is written in C\#, the new algorithms are (necessarily) written in C, called from the main tool via the P/Invoke mechanism.
We used GCC 10.3.0 to compile our implementations on both 64-bit Linux and Windows~10.
We manually inspected the disassembly of the generated code to ensure that GCC's optimisations did not interfere with rounding mode changes as described in \Cref{sec:RoundingInLanguages}.
In a significant architectural change, we modified \mcsta's state space exploration and representation code to preserve the exact rational values for the probabilities specified in the model, so that safely-rounded floating-point representations for the $\mu(s')$ can be computed during iteration as described above.


Of each algorithm, we implemented four variants:
a \textit{default} one that leaves the choice of instruction set to the compiler and uses \texttt{fesetround} to change rounding mode;
an \textit{x87} variant that forces floating-point operations to use the x87 instructions by attributing the relevant functions with \texttt{target("fpmath=387")} and that changes rounding mode via inline assembly using FNSTCW/FLDCW;
an \textit{SSE} variant that forces the SSE instruction set via \texttt{target("fpmath=sse")} and uses VSTMXCSR/VLDMXCSR in inline assembly for rounding mode chan\-ges; and
an \textit{AVX-512} variant that implements all floating-point operations requiring non-default rounding modes via AVX-512 intrinsics, in particular using \texttt{\_mm\_fmadd\_round\_sd} in the Bellman equations.
All variants use double precision; \textit{default} and \textit{SSE} additionally have a single-precision version (which we omit for \textit{x87} since the reduced precision does not speed up the operations we use); and \textit{x87} also provides an 80-bit extended-precision version (however we currently return its results as safely-rounded double-precision values due to the unavailability of a \texttt{long} \texttt{double} equivalent in C\#, which limits its use outside of performance testing for now).
All in all, we thus provide 28 variants of interval iteration for comparison, out of which 14 provide guaranteed correct results.

In particular, the safe rounding makes PMC feasible at 32-bit single precision, which would otherwise be too likely to produce incorrect results.
While we expect that this may deliver many results with low precision (but which are correct) due to a rounded fixpoint being reached long before the relative width reaches $\epsilon$, it also halves the memory needed to store $l$ and $u$, and may speed up computations.
At the opposite end, \mcsta is now also the first PMC tool that can use 80-bit extended precision, which however doubles the memory needed for $l$ and $u$ since 80-bit \texttt{long} \texttt{double} values occupy 16 bytes in memory (with GCC).

\section{Experiments}
\label{sec:Experiments}

Using our implementation in \mcsta, we first tested all variants of the algorithms on $M_n^\gamma$ in the setting of \Cref{sec:ErrorsInPMC}.
As expected, and validating the correctness of the approach and its implementation, all SR variants return \textit{unknown}.

We then assembled a set of 31 benchmark instances---combinations of a model, values for its configurable parameters, and a property to check---from the QVBS covering DTMC, MDP, and probabilistic timed automata (PTA)~\cite{KNSS02} transformed to MDP by \mcsta using the digital clocks approach~\cite{KNPS06}.
These are all the models and probabilistic reachability probabilities from the QVBS supported by \mcsta for which the result was not $0$ or $1$ (then it can be computed via graph-based algorithms) and for which a parameter configuration was available where PMC terminated within our timeout of 120\,s but II needed enough time for it to be measured reliably ($\gtrapprox 0.2$\,s).
We checked each of these benchmarks with all 28 variants of our algorithms on the following x86-64 systems:
\textbf{I11w}: an Intel Core i5-1135G7 (up to $4.2$\,GHz) laptop running Windows~10, this being the only system we had access to with AVX-512 support;
\textbf{AMDw}: an AMD Ryzen~9 5900X ($3.7$-$4.8$\,GHz) workstation running Windows~10, representing current AMD CPUs in our evaluation;
\textbf{I4x}: an Intel Core i7-4790 ($3.6$-$4.0$\,GHz) workstation running Ubuntu Linux 18.04, representing older-generation Intel desktop hardware; and
\textbf{IPx}: an Intel Pentium Silver J5005 ($1.5$-$2.8$\,GHz) compact PC running Ubuntu Linux 18.04, representing a non-Core low-power Intel system. 
We show a selection of our experimental results in the remainder of this section,
mainly from I11w and AMDw.
We remark on cases where the other systems (all with Intel CPUs) showed different patterns from I11w.

\begin{figure}[t]
\centering
\input{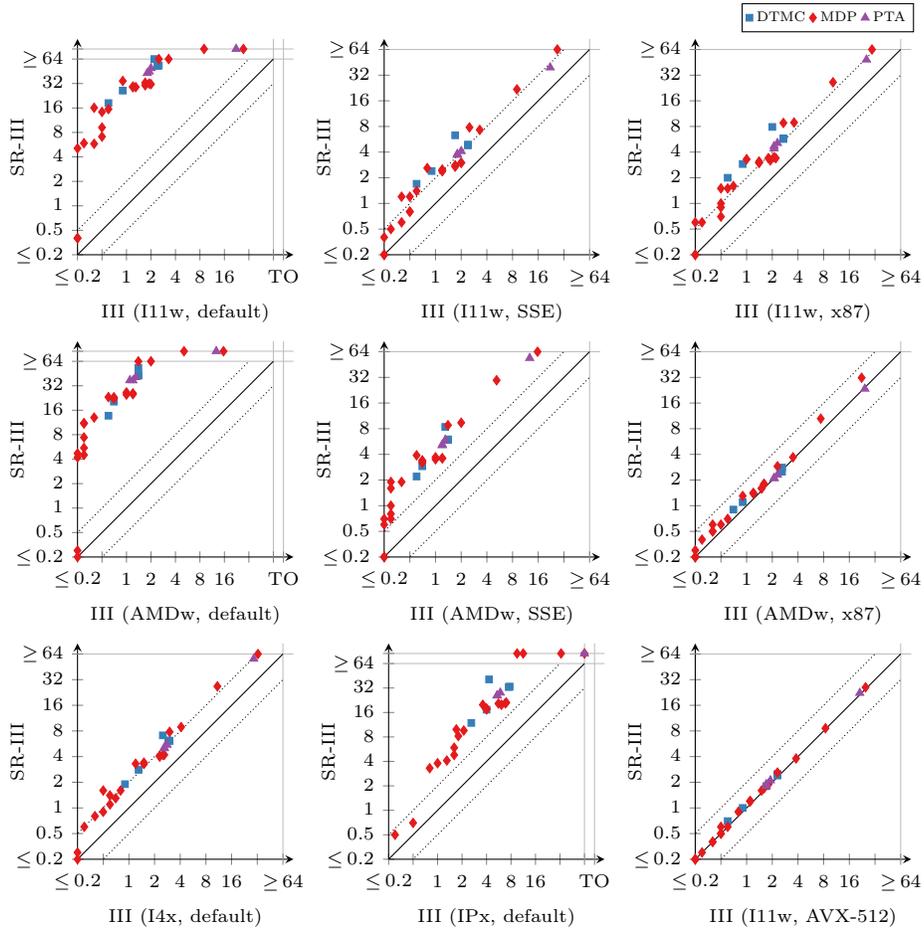}
\caption{Performance impact of safe rounding across instruction sets and systems}
\label{fig:ScatterDefaultVsSafe}
\end{figure}

We present results graphically as scatter plots like in \Cref{fig:ScatterDefaultVsSafe}.
Each such plot compares two algorithm variants in terms of runtime for the iteration phase of the algorithm only (\ie we exclude the time for state space exploration and preprocessing).
Every point $\tuple{x, y}$ corresponds to a benchmark instance and indicates that the variant noted on the x-axis took $x$ seconds to solve this instance while the one noted on the y-axis took $y$ seconds.
Thus points above the solid diagonal line correspond to instances where the x-axis method was faster; points above (below) the upper (lower) dotted diagonal line are where the x-axis method took less than half (more than twice) as long.

%

\Cref{fig:ScatterDefaultVsSafe} first shows the performance impact of enabling safe rounding for the standard interleaved algorithm using double precision.
The top row shows the behaviour on I11w.
We see that runtime is drastically longer in the \textit{default} variant that uses \texttt{fesetround}, but only increases by a factor of around 2 if we use the specific inline assembly instructions.
We note that GCC includes the code for \texttt{fesetround} in the generated \texttt{.dll} file on Windows, but in contrast to the assembly methods does not inline it into the callers.
Some of the difference may thus be function call overhead.
The middle row shows the behaviour on AMDw.
Here, \textit{default} is affected just as badly, but the effect on \textit{SEE} is worse while that on \textit{x87} is much lower than on the Intel I11w system.
In the bottom row, we show the impact on \textit{default} on the Linux systems (bottom left and bottom middle), which is much lower than on Windows.
This is despite GCC implementing \texttt{fesetround} as an external library call here.
The overhead still markedly differs between the two Intel CPUs, though.
Finally, as expected, we see on the bottom right than safe rounding has almost no performance impact when using the AVX-512 instructions.

%
\begin{figure}[t]
\centering
\input{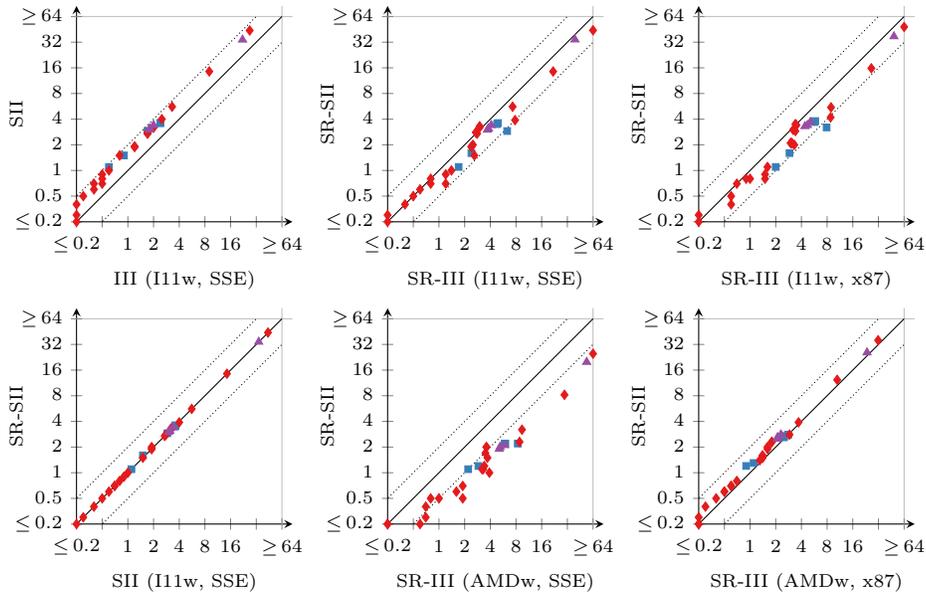}
\caption{Performance of interleaved compared to sequential II}
\label{fig:ScatterInterleavedVsSequential}
\end{figure}

Seeing the significant impact that enabling safe rounding can have, we next show what the sequential algorithm brings to the table, in \Cref{fig:ScatterInterleavedVsSequential}.
On the top left, we compare the base algorithms without safe rounding, where SII takes up to twice as long in the worst case.
This is likely due to the more cache-friendly memory access pattern of III:
we store $l$ and $u$ interleaved for III, so it always operates on two adjacent values at a time.
The bottom-left plot confirms that reducing the number of rounding mode changes reduces the overhead of safe rounding to essentially zero.
The remaining four plots show the differences between SR-III and SR-SII.
In all cases except \textit{x87} on AMDw, SR-III is slower.
We thus have that III is fastest but unsafe, SII and SR-SII are equally fast but the latter is safe, and SR-III is safe but tends to be slower on the Intel systems.
On the AMD system, SR-III surprisingly wins over SR-SII with \textit{x87}, highlighting that the x87 instruction set in Ryzen~3 must be implemented very differently from SSE.

%
\begin{figure}[t]
\centering
\input{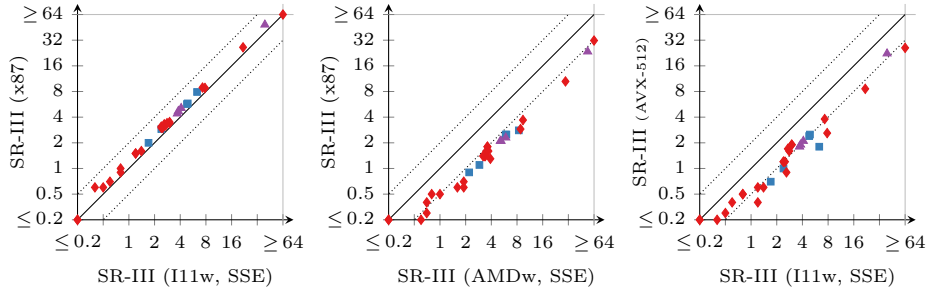}
\caption{Performance with different instruction sets}
\label{fig:ScatterInstructionSet}
\end{figure}

We further investigate the impact of the instruction set in \Cref{fig:ScatterInstructionSet}.
Confirming the patterns we saw so far, SSE is slightly faster than x87 on I11w (and we see similar behaviour on the other Intel systems) but slower by a factor of more than 2 on the AMD CPU.
The rightmost plot highlights that AVX-512 is the fastest alternative on the most recent Intel CPUs, which may in part be due to the availability of the fused multiply-add instruction that fits II so well.

%
\begin{figure}[t]
\centering
\input{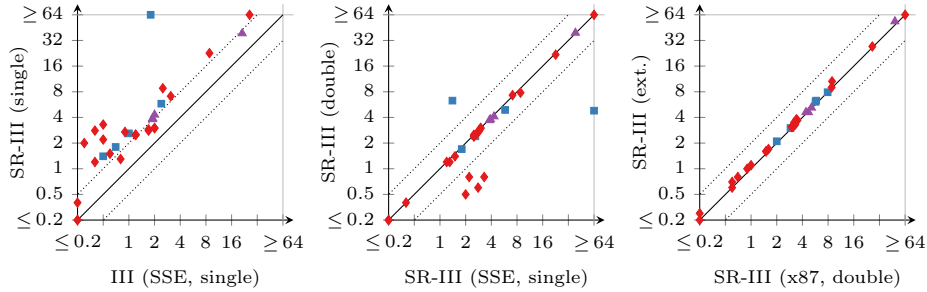}
\caption{Performance with different precision settings (on I11w)}
\label{fig:ScatterPrecision}
\end{figure}

All results so far were for double-precision computations.
To conclude our evaluation, we show in \Cref{fig:ScatterPrecision} that reducing to single precision does not bring the expected performance benefits.
We see in the leftmost plot that the overhead of safe rounding has a much higher variance compared to \Cref{fig:ScatterDefaultVsSafe}.
The detailed tool outputs hint at the reason being that rounding away from the fixpoint occurs in much larger steps with single precision, which significantly slows down or stops the convergence in several instances.
The middle plot shows that, aside from the slowly converging outliers, using single precision does not provide a speedup over using doubles.
Finally, on the right, we show that the impact of enabling 80-bit extended precision on x87 is minimal.

\section{Conclusion}
\label{sec:Conclusion}

There has been ample research into sound PMC algorithms over the past years, but the problem of errors introduced by naive implementations using default floating-point rounding has been all but ignored.
We showed that a solution exists that, while perhaps conceptually simple, faces a number of implementation and performance obstacles.
In particular, hardware support for rounding modes is arguably essential to achieve acceptable performance, but difficult to use from C/C++ and impossible to access from most other programming languages.
We extensively explored the space of implementation variants, highlighting that performance crucially depends on the combination of the variant, the CPU, and the operating system.
Nevertheless, our results show that truly correct PMC is possible today at a small cost in performance, which should all but disappear as AVX-512 is more widely adopted.
With our implementation in \mcsta, we provide the first PMC tool that combines fast, scalable, \emph{and} correct.


At this point, the only remaining cause for incorrect results are implementation bugs.
However, our paper also paves the way towards formally verified correct-by-construction implementations of PMC using \eg the Isabelle refinement framework~\cite{Lam19}:
with default rounding, we could never prove correctness because the algorithm \emph{is not} correct.


\paragraph{Acknowledgments.}
This work was triggered by Masahide Kashiwagi's excellent overview of the different ways to change rounding mode as used by his \texttt{kv} library for verified numerical computations~\cite{kvlib}.
The author thanks Anke and Ursula Hartmanns for contributing to the diversity of hardware on which the experiments were performed by providing access to the AMDw and I11w systems.


\bibliography{paper}
\bibliographystyle{splncs04}

\end{document}